\documentclass{elsart}
\usepackage{hyperref}
\usepackage[square,comma,sort&compress]{natbib}
\usepackage{graphicx}
\usepackage{hyphenat}
\usepackage{psfrag}
\usepackage[latin1]{inputenc}
\usepackage{amssymb}
\newlength{\sepmod}
\newcommand{\ds}{\displaystyle}

\newcommand{\tD}{\tilde{\Delta}}
\newcommand{\td}{\tilde{d}}

\begin{document}

\begin{frontmatter}

\title{Geometric and Stochastic Clusters of Gravitating Potts Models}
%\author{Martin Weigel\thanksref{mw} and Wolfhard Janke\thanksref{wj}}
%\thanks[wj]{janke@itp.uni-leipzig.de}

\author{Wolfhard Janke}
\ead{Wolfhard.Janke@itp.uni-leipzig.de}
\address{Institut f\"ur Theoretische Physik and Centre for Theoretical Sciences
  (NTZ), Universit\"at Leipzig, Augustusplatz
  10/11, 04109~Leipzig, Germany}
\and
\author{Martin Weigel}
\ead{M.Weigel@ma.hw.ac.uk}
\address{Department of Mathematics, Heriot-Watt University, Riccarton, Edinburgh,
  EH14\,4AS, Scotland}

\begin{abstract}
  We consider the fractal dimensions of critical clusters occurring in configurations
  of a $q$-state Potts model coupled to the planar random graphs of the dynamical
  triangulations formulation of Euclidean quantum gravity in two dimensions. For
  regular lattices, it is well-established that at criticality the properties of
  Fortuin-Kasteleyn clusters are directly related to the conventional critical
  exponents, whereas the corresponding properties of the geometric clusters of like
  spins are not. Recently it has been observed that the latter are
  related to the critical properties of a tricritical Potts model with the same
  central charge.  We apply the KPZ formalism to develop a related prediction for the
  case of Potts models coupled to quantum gravity and employ numerical simulation
  methods to confirm it for the Ising case $q=2$.
\end{abstract}


\end{frontmatter}

\section{Introduction}

The relation between the percolation problem and thermal phase transitions of lattice
spin systems has been a question of intense research for at least three decades.
Clusters of even spins are natural objects occurring in the analysis of phase
ordering processes and nucleation \cite{binder:76}, and a theory of critical
phenomena in terms of purely geometrical objects appears appealing. In this context,
it had long been surmised that a continuous phase transition of a spin system might
be accompanied (or, in fact, caused) by a percolation transition of the clusters of
like spins ({\em geometric clusters\/}), the appearance of a percolating cluster
sustaining the onset of a non-zero magnetisation \cite{bishop:74}. While for the
special case of the Potts model in two dimensions it turned out that, indeed, the
thermal phase transition point coincides with the percolation transition of the spin
clusters, this behaviour is not generic and does not occur in three-dimensional
systems \cite{coniglio:77}. Also, the critical exponents associated to the
percolation of geometric clusters are not directly related to the thermal exponents
of the spin model \cite{sykes:76}. However, a close relation between the percolation
and thermal phase transitions can be established by considering {\em
  stochastically\/} defined clusters (or droplets) as they occur in the
Fortuin-Kasteleyn (FK) representation of the Potts model, and it can be shown that,
in fact, the Potts model is equivalent to a site-bond correlated percolation problem
\cite{fortuinandco} such that the corresponding critical exponents agree. This
identification of the proper cluster objects ({\em FK clusters\/}) percolating at the
thermal phase transition subsequently also allowed for the design of {\em cluster
  algorithms\/} for the efficient simulation of Potts models in the vicinity of the
ordering transition \cite{swendsen-wang:87a,wolff:89a}, beating the observed critical
slowing down of local update algorithms. Similarly, relations of continuous-spin
models to percolation problems could be established and corresponding cluster
algorithms formulated \cite{wolff:89a,blanchard:00}, such that the continuous phase
transitions %\footnote{\tt What about percolation and {\em first-order\/} transitions?}
of many standard models of statistical mechanics are by now understood in terms of
the percolation properties of some suitably defined ensemble of stochastic clusters.

Although not in the universality class of the thermal phase transition, the clusters
of like spins or geometric clusters still undergo a percolation transition in the
course of thermal phase ordering. This transition is in general not equivalent to
ordinary (site or bond) percolation \cite{sykes:76} and it remains an interesting
open question to determine the general critical behaviour of clusters of aligned
spins. For the case of the two-dimensional Ising model, it has been conjectured and
numerically verified that the geometric clusters are described by the $q=1$
tricritical Potts model \cite{vanderzande:89}; this correspondence can be understood
from a direct construction starting from the dilute Potts model \cite{stella:89}.
Subsequently, analogous conjectures for the $2 < q \le 4$ Potts models were made and
some of them substantiated by numerical simulations \cite{duplantier}. Analytical
calculations concerning clusters occurring in systems of statistical physics and
their boundaries, traditionally based on methods of conformal field theory and the
Coulomb gas \cite{duplantier:03}, have recently seen major advances from the insight
that fractal random curves can be described in a framework dubbed {\em stochastic
  Loewner evolution\/} (SLE) \cite{schramm:00,bauer:06}. Collecting these
observations, a more systematic analysis of the relation between the critical and
tricritical branches of the Potts model and their connection to the FK and geometric
clusters has been performed, resulting in the identification of exact values for the
different cluster fractal dimensions and their numerical verification
\cite{deng,janke}.

%\begin{itemize}
%\item {\tt mention further Duplantier articles\/}
%\item SLE only relates to {\em hulls\/}
%\end{itemize}

Coupling spin models to the planar random lattices of the dynamical triangulations
model of two-dimensional Euclidean quantum gravity \cite{ambjorn:book} corresponds to
the introduction of a particular type of annealed connectivity disorder. The
resulting randomness is strong, leading to a change in critical behaviour of virtually
all types of coupled matter variables \cite{kpz}.  Geometrically, it is characterised
by a large fractal dimension $d_h \approx 4$ of these lattices of topological
dimension two. It is interesting to see how the relation between geometric and FK
clusters of the Potts model works out far away from the regularity of a Bravais
lattice. The mentioned relations between geometric and FK clusters have not yet been
rigorously established, such that evidence from further models is highly welcome
support. Besides, the behaviour of these fractal cluster objects on lattices which are
themselves highly fractal is of particular interest in itself.

\section{Dynamical triangulations and the KPZ formula}

The dynamical triangulations approach provides a constructive model for Euclidean
quantum gravity in general dimensions \cite{ambjorn:book}. It regularizes the path
integral over fluctuating metrics naturally occurring in an attempt to quantize the
gravitational interaction by a sum over combinatorial manifolds realised as
simplicial complexes \cite{ambjorn:book2}. In two dimensions, the resulting canonical
ensemble of discrete surfaces can be defined as that of all possible gluings of a
given number $N_2$ of equilateral triangles to a closed surface of fixed (usually
spherical) topology, where all resulting triangulations are counted with the same
weight in the partition sum. This is a purely combinatorial definition, and due to
the equilaterality of the triangles the resulting triangulations cannot in general be
embedded in the Euclidean $\mathbb{R}^3$ and, in fact, one is only interested in
their {\em intrinsic\/} geometry. Many counting problems related to these graphs can
be solved exactly by using matrix integrals or general combinatorial techniques
\cite{tutte}. For our purposes it is sufficient to note that the resulting random
graphs are highly fractal with blobs (``baby universes'') of arbitrary size being
connected to the main graph body with a minimal number of links.  This structure
entails an internal Hausdorff dimension $d_h = 4$ \cite{ambjorn:book}.  Decorating
the vertices (or edges, faces) of the graphs with matter variables, corresponding to
an annealed geometrical average, leads to a change of universality class at
criticality, expressed through a dressing of the conformal weights $\Delta$ of the
matter fields given by the KPZ formula \cite{kpz},
\begin{equation}
  \tilde{\Delta} = \frac{\sqrt{1-c+24\Delta}-\sqrt{1-c}}{\sqrt{25-c}-\sqrt{1-c}},
  \label{dressing_KPZ}
\end{equation}
where $c$ denotes the central charge of the coupled matter model. Analogously, the
change of the string-susceptibility exponent $\gamma_s$ can be expressed in terms of
$c$ \cite{kpz}.

As mentioned above, the fractal properties of FK and geometric clusters of the
square-lattice Ising model have been studied rather extensively
\cite{stella:89,vanderzande:89,janke}. Here, we are interested in the (normalised)
fractal dimensions $d_\mathrm{C}^\mathrm{FK}/d$ resp.\ $d_\mathrm{C}^\mathrm{G}/d$ of
the incipient percolating Fortuin-Kasteleyn resp.\ geometric cluster at criticality
(where $d=2$ denotes the spatial dimension). For the FK clusters,
$d_\mathrm{C}^\mathrm{FK}/d = 1-\beta/d\nu = 15/16$ is an exact result
\cite{coniglio:80a}. From the identification of the geometric Ising clusters with the
FK clusters of a $q=1$ tricritical Potts model, one finds $d_\mathrm{C}^\mathrm{G}/d
= 187/192$ \cite{stella:89}.  These values are related to the conformal weights
$\Delta$ above as $\Delta_\mathrm{C} = 1-d_\mathrm{C}/d $ \cite{henkel:book}. To
describe the coupling of the two-dimensional Ising model to quantum gravity, the
resulting weights $\Delta_C^\mathrm{FK} = 1/16$ and $\Delta_C^\mathrm{G} = 5/192$
have to be dressed according to Eq.~(\ref{dressing_KPZ}). Since for both, the Ising
and $q=1$ tricritical Potts models, the central charge $c = 1/2$, from
(\ref{dressing_KPZ}) we find $\tilde{\Delta}_C^\mathrm{FK} = 1/6$ and
$\tilde{\Delta}_C^\mathrm{G} = 1/12$.  Therefore, we expect the following cluster
fractal dimensions for the Ising model coupled to dynamical triangulations,
\begin{equation}
  \label{eq:conjecture}
  \tilde{d}_\mathrm{C}^\mathrm{FK}/d_h = 1-\tilde{\Delta}_C^\mathrm{FK} = 5/6, \hspace{0.5cm}
  \tilde{d}_\mathrm{C}^\mathrm{G}/d_h = 1-\tilde{\Delta}_C^\mathrm{G} = 11/12,
\end{equation}
which now have been written in units of the Hausdorff dimension $d_h$ of the graphs
coupled to the spin model, the value of which is not known exactly but numerically
found to be consistent with $d_h = 4$ for all $0\le c\le 1$
\cite{ambjorn:95d,weigel:04b}.

\section{Numerical simulation method and results for the Ising model}

For a numerical simulation of the system, due to the annealed nature of the disorder
both, the underlying dynamical triangulations as well as the coupled Ising spins,
must be updated in parallel. An ergodic set of Monte Carlo updates for the dynamical
triangulations is given by the so-called Pachner moves \cite{pachner:91a}. For the
canonical ensemble of a fixed number $N_2$ of triangles in two dimensions, these
reduce to the following flip between two adjacent triangles,
\begin{center}
  \psfrag{p1}{$p_1$}
  \psfrag{p2}{$p_2$}
  \psfrag{p3}{$p_3$}
  \psfrag{q1}{$q_1$}
  \psfrag{q2}{$q_2$}
  \begin{tabular}{ccc}
    \includegraphics[clip=true,keepaspectratio=true,width=2.75cm]{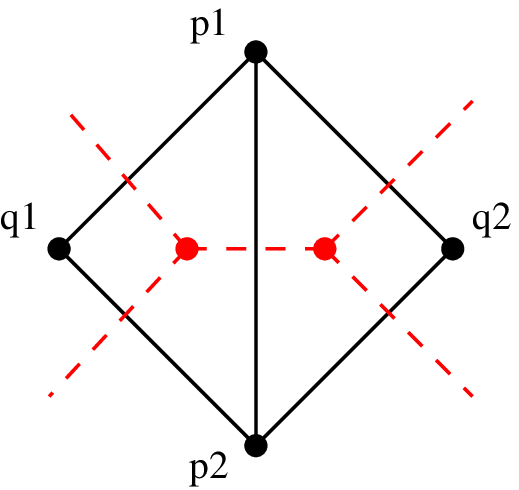} &
    \raisebox{1.2cm}{\includegraphics[clip=true,keepaspectratio=true,width=1.5cm]{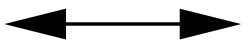}} &
    \includegraphics[clip=true,keepaspectratio=true,width=2.75cm]{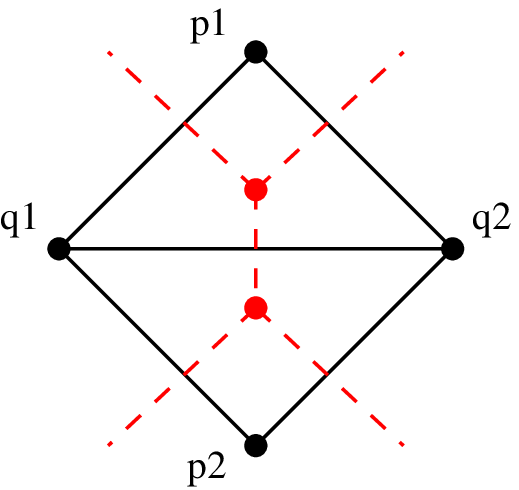}
  \end{tabular}  
\end{center}
where the dashed lines indicate the effect on the dual $\phi^3$ graph. To improve the
decorrelation of adjacent configurations in the Monte-Carlo Markov chain, this local
update is complemented by intermittent non-local rewirings of the blob structure of
the graph known as ``baby-universe surgery method'' \cite{ambjorn:94a,doktor}. We
consider the case of non-degenerate triangulations, corresponding to dual one-point
irreducible $\phi^3$ Feynman diagrams and place the Ising spins on the faces of the
triangulation or, equivalently, the vertices of the dual $\phi^3$ graphs. The coupled
Ising model is being updated according to the Swendsen-Wang cluster algorithm
\cite{swendsen-wang:87a} in order to alleviate the critical slowing down of the spin
variables expected since simulations are performed at the exactly known inverse
critical temperature $\beta_c = \frac{1}{2}\ln \frac{108}{23} \approx 0.7733$ of the
coupled system \cite{burda:89a}.

Although the fractal dimensions $d_\mathrm{C}^\mathrm{FK}$ and
$d_\mathrm{C}^\mathrm{G}$ could be determined directly by means of an appropriate
geometrical analysis of the clusters, it is convenient to exploit relations to more
easily accessible quantities. The fractal dimension of Fortuin-Kasteleyn clusters,
$d_\mathrm{C}^\mathrm{FK}$, is related to the magnetic susceptibility exponent
$\gamma = \gamma^\mathrm{FK}$ as $d_\mathrm{C}/d_h = \frac{1}{2}(1+\gamma/d_h\nu)$
\cite{stauffer:book}, where $\nu$ denotes the critical exponent of the correlation
length. The susceptibility $\chi = \chi^\mathrm{FK}$ can be sampled by the following
estimator in the cluster language \cite{wolff:88a},
\begin{equation}
  \label{eq:improved_est}
  \chi \equiv \frac{1}{N_2} \left\langle \left( \sum_i \sigma_i \right)^2 \right\rangle
       = \frac{1}{N_2} \left\langle \sum_i |C_i|^2 \right\rangle,
\end{equation}
where $\sigma_i = \pm 1$ denotes the Ising spins. The sum in the second expression is
over all FK clusters $C_i$ (including the percolating cluster as well as isolated
sites) of a given configuration and $|C_i|$ denotes the number of spins in cluster
$i$. Close to the critical point and for a system of finite size $N_2$, standard
finite-size scaling (FSS) arguments suggest that $\chi$ scales as
\begin{equation}
  \label{eq:FSS}
  \chi \sim N_2^{\gamma/d_h\nu} = N_2^{2d_\mathrm{C}/d_h-1}.
\end{equation}
Thus, Eqs.~(\ref{eq:improved_est}) and (\ref{eq:FSS}) allow for a FSS determination
of the fractal dimension $d_\mathrm{C}^\mathrm{FK}/d_h$ of the FK clusters from
information about the cluster distribution naturally produced by the Swendsen-Wang
cluster update of the spins. In complete analogy, a ``geometric susceptibility''
$\chi^\mathrm{G}$ can be estimated with $C_i$ now denoting the {\em geometric\/}
clusters in relation (\ref{eq:improved_est}), and its FSS is described by
Eq.~(\ref{eq:FSS}) with exponent $d_\mathrm{C}^\mathrm{G}/d_h$.

\begin{figure}[tb]
  \centering
  \includegraphics[clip=true,keepaspectratio=true,width=11cm]{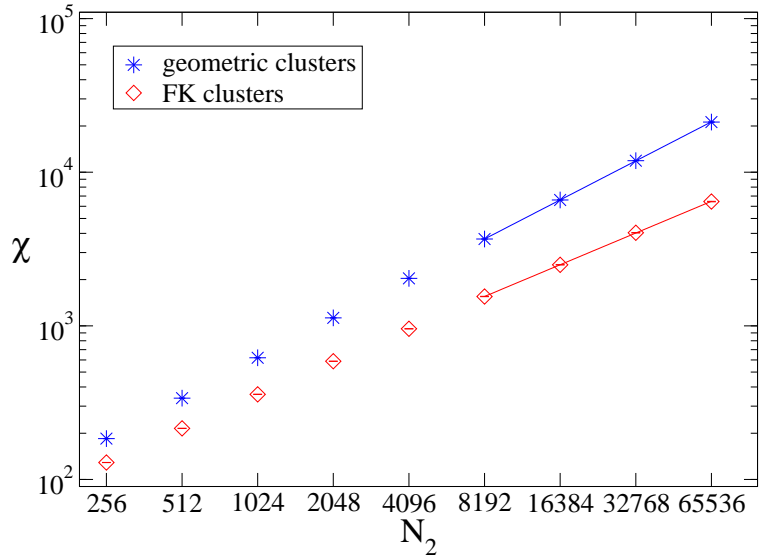}
  \caption
  {Finite-size scaling of the magnetic (``FK clusters'') and geometric (``geometric
    clusters'') susceptibilities $\chi$ of the Ising model coupled to regular
    dynamical triangulations with $N_2$ triangles and at the critical inverse
    temperature $\beta_c = \frac{1}{2}\ln \frac{108}{23} \approx 0.7733$. The lines
    show fits of the functional form (\ref{eq:FSS}) to the data.}
  \label{fig:scaling}
\end{figure}

As a check and gauge for the method, we determined the cluster fractal dimensions for
the Ising model on the square lattice, using $L\times L$ lattices with
$L=8,\ldots,512$. From fits of the form (\ref{eq:FSS}) to the (magnetic and
geometric) susceptibility data, we find $d_\mathrm{C}^\mathrm{FK}/d = 0.93\,756(15)$
and $d_\mathrm{C}^\mathrm{G}/d = 0.97\,370(13)$ (including lattices of sizes
$L=16,\ldots,512$), in very good agreement with the exact results
$d_\mathrm{C}^\mathrm{FK}/d = 15/16 \approx 0.93\,750$ and $d_\mathrm{C}^\mathrm{G}/d
= 187/192 \approx 0.97\,396$. For the dynamical triangulations model, only graphs of
somewhat smaller sizes can be equilibrated, and we consider volumes $N_2 =
256,512,\ldots,65\,536$.  Figure \ref{fig:scaling} shows the resulting susceptibility
data in a doubly logarithmic plot together with fits of the form (\ref{eq:FSS}) to
the data. Due to rather strong finite-size corrections, only the largest lattice
sizes can be included in the uncorrected fit (\ref{eq:FSS}) with reasonable fit
quality. For the range $N_2=8192,\ldots,65536$, we arrive at the estimates
$\tilde{d}_\mathrm{C}^\mathrm{FK}/d_h = 0.84\,301(93)$ and
$\tilde{d}_\mathrm{C}^\mathrm{G}/d_h = 0.92\,220(55)$, in marginal agreement with the
conjectured values of Eq.~(\ref{eq:conjecture}),
$\tilde{d}_\mathrm{C}^\mathrm{FK}/d_h \approx 0.83\,333$ and
$\tilde{d}_\mathrm{C}^\mathrm{G}/d_h \approx 0.91\,667$. From previous experience
with the dynamical triangulations model, one indeed expects very strong finite-size
corrections to be present \cite{weigel:04b}.  These result from the small effective
linear extents of lattices of a given number of triangles reflected in the large
fractal dimension $d_h \approx 4$. As dominant contribution, we expect analytic
corrections to the effective linear extent,
\begin{equation}
  \label{eq:leff}
  L_\mathrm{eff}(N_2) = L_0 N_2^{1/d_h}+ L_1 + L_2 N_2^{-1/d_h}+\cdots,
\end{equation}
such that the susceptibility should scale as $\chi \sim
[L_\mathrm{eff}(N_2)]^{\gamma/\nu}$. Since the data are not precise enough to
determine an additional exponent, we fix $d_h = 4$ in the expression for
$L_\mathrm{eff}(N_2)$. Already with one correction term only ($L_2 = 0$), the fit
results,
\begin{equation}
  \tilde{d}_\mathrm{C}^\mathrm{FK}/d_h = 0.8291(19),\;\;\; \tilde{d}_\mathrm{C}^\mathrm{G}/d_h
  = 0.9132(12)
\end{equation}
for the range $N_2=1024,\ldots,65536$, move considerably closer to the conjectured
values. With both correction terms (i.e., variable $L_2$), we get full consistency
with estimates $\tilde{d}_\mathrm{C}^\mathrm{FK}/d_h = 0.8349(118)$ and
$\tilde{d}_\mathrm{C}^\mathrm{G}/d_h = 0.9164(68)$. Conversely, if we {\em fix\/}
$\tilde{d}_\mathrm{C}^\mathrm{FK}/d_h$ resp.\ $\tilde{d}_\mathrm{C}^\mathrm{G}/d_h$
at the conjectured values, the fits with one correction term give reasonable fit
qualities with $Q=0.20$ resp.\ $Q=0.03$, and very high-quality fits are reached when
including both correction terms with $Q=0.66$ resp.\ $Q=0.42$.

\section{General results for the Potts model}

To understand the general relation between FK and geometric clusters and the critical
and tricritical branches of the Potts model coupled to dynamical triangulations, we
consider the following parametrisation of the critical $q$-state Potts model
\cite{salas:97a},
\begin{equation}
  \sqrt{q} = -2\cos(\pi/\kappa),\;\;\;\kappa = \frac{1+m}{m},\;\;\;m=1,2,3,\ldots,
  \label{eq:critical}
\end{equation}
where $1\le\kappa\le 2$ and the central charge $c = 1-6/m(m+1)$, such that
$\kappa=2$, $3/2$, $4/3$, $6/5$, $1$ corresponds to the $q=0$, $1$, $2$, $3$, $4$
Potts models with $c=-2$, $0$, $1/2$, $4/5$, $1$, respectively. The conformal weights
of the primary operators follow from the Kac table \cite{henkel:book},
\begin{equation}
  \Delta_{r,s} = \frac{[(m+1)r-ms]^2-1}{4m(m+1)},\;\;\;r,s = 1,2,\ldots,m-1,
  \label{eq:kactable}
\end{equation}
and they are identified with the physical operators of the Potts model as follows,
\begin{equation}
  \begin{array}{rcccl}
    \Delta_{\epsilon} & = & \Delta_{2,1} & = & \ds\frac{3\kappa}{4}-\frac{1}{2},\\[1.5ex]
    \Delta_{\sigma} & = & \Delta_{\frac{m+1}{2}-1,\frac{m+1}{2}} & = &
    \ds\frac{1}{2}-\frac{3\kappa}{16}-\frac{1}{4\kappa},\\[1.5ex]
    \Delta_\mathrm{C} & = & \Delta_{\frac{m+1}{2},\frac{m+1}{2}}
    & = & \ds\frac{1}{2}-\frac{3\kappa}{16}-\frac{1}{4\kappa},\\[1.5ex]
    \Delta_\mathrm{H} & = & \ds\Delta_{m,m} & = & \ds\frac{1}{2}-\frac{\kappa}{4},\\[1.5ex]
    \Delta_\mathrm{EP} & = & \ds\Delta_{m+1,m+1} & = & \ds\frac{1}{2}-\frac{1}{4\kappa},\\[1.5ex]
    \Delta_\mathrm{RB} & = & \ds\Delta_{m,m-1} & = & \ds\frac{1}{2}+\frac{3}{4\kappa}-\frac{\kappa}{4},
  \end{array}
  \label{eq:critical_operators}
\end{equation}
where $\Delta_{\epsilon}$ and $\Delta_{\sigma}=\Delta_\mathrm{C}$ correspond to the
leading energetic and magnetic operators, respectively, $\Delta_\mathrm{H}$ denotes
the weight corresponding to the cluster hull, $\Delta_\mathrm{EP}$ the weight of the
external perimeter and $\Delta_\mathrm{RB}$ the weight of the ``red bonds'' of the
cluster, cf.\ Ref.~\cite{janke}. The various fractal dimensions of FK clusters are
given by the corresponding renormalization-group eigenvalues, $d_\alpha = y_\alpha
\equiv d_h(1-\Delta_\alpha)$, where $\alpha=\mathrm{C}$, $\mathrm{H}$, $\mathrm{EP}$,
$\mathrm{RB}$.

In Ref.~\cite{janke} it was argued that in general the fractal dimensions of the
geometric clusters of the critical Potts model are identical to the fractal
dimensions of the {\em tricritical\/} Potts model of the same central charge, to be
reached by performing the central-charge conserving ``duality transformation''
$\kappa \rightarrow 1/\kappa$ in Eq.~(\ref{eq:critical}). For the tricritical model,
the leading conformal weights are identified as
\begin{equation}
  \begin{array}{rcccl}
    \Delta_{\epsilon} & = & \Delta_{1,2} & = & \ds\frac{3}{4\kappa}-\frac{1}{2},\\[1.5ex]
    \Delta_{\sigma} & = & \Delta_{\frac{m}{2},\frac{m}{2}} & = &
    \ds\frac{1}{2}-\frac{3}{16\kappa}-\frac{\kappa}{4},\\[1.5ex]
    \Delta_\mathrm{C} & = & \Delta_{\frac{m}{2},\frac{m}{2}}
    & = & \ds\frac{1}{2}-\frac{3}{16\kappa}-\frac{\kappa}{4},\\[1.5ex]
    \Delta_\mathrm{H} & = & \ds\Delta_{m+1,m+1} & = & \ds\frac{1}{2}-\frac{1}{4\kappa},\\[1.5ex]
    \Delta_\mathrm{EP} & = & \ds\Delta_{m,m} & = & \ds\frac{1}{2}-\frac{\kappa}{4},\\[1.5ex]
    \Delta_\mathrm{RB} & = & \ds\Delta_{\frac{m+4}{2},\frac{m+1}{2}} & = &
    \ds\frac{1}{2}+\frac{3\kappa}{4}-\frac{1}{4\kappa},
  \end{array}
  \label{eq:tricritical_operators}
\end{equation}
which, as can be seen, directly follow from those of
Eq.~(\ref{eq:critical_operators}) by replacing $\kappa \rightarrow 1/\kappa$.

For the weights of the unitary minimal models of Eq.~(\ref{eq:kactable}), the KPZ
formula (\ref{dressing_KPZ}) can be written as
\begin{equation}
  \label{eq:kpz_minimal_explicit}
  \tilde{\Delta}_{r,s} = \frac{1}{2}(1-\kappa+|s-\kappa r|),
\end{equation}
explicitly revealing that all weights of the minimal models dressed for the coupling
to quantum gravity are rational numbers. Dressing the weights (\ref{eq:critical_operators}) of
the {\em critical branch\/} yields
\begin{equation}
  \label{eq:dressed_critical}
  \begin{array}{rclcrcl}
    \tD_{\epsilon} & = & \ds\frac{\kappa}{2}, & & \tD_{\sigma} & = &
    \ds\frac{1}{2}-\frac{\kappa}{4}, \\[1.5ex]
    \tD_\mathrm{C} & = & \ds\frac{1}{2}-\frac{\kappa}{4}, & & \tD_\mathrm{H} & = &
    \ds 1-\frac{\kappa}{2}, \\[1.5ex]
    \tD_\mathrm{EP} & = & \ds\frac{1}{2}, & & \tD_\mathrm{RB} & = & \ds\frac{3}{2}-\frac{\kappa}{2},
  \end{array}
\end{equation}
whereas the weights (\ref{eq:tricritical_operators}) of the {\em tricritical
  branch\/} become
\begin{equation}
  \label{eq:dressed_tricritical}
  \begin{array}{rclcrcl}
    \tD_{\epsilon} & = & \ds\frac{3}{2}-\kappa, & & \tD_{\sigma} & = &
    \ds\frac{3}{4}-\frac{\kappa}{2}, \\[1.5ex]
    \tD_\mathrm{C} & = & \ds\frac{3}{4}-\frac{\kappa}{2}, & & \tD_\mathrm{H} & = &
    \ds \frac{1}{2}, \\[1.5ex]
    \tD_\mathrm{EP} & = & \ds1-\frac{\kappa}{2}, & & \tD_\mathrm{RB} & = & \ds\frac{1}{2}+\frac{\kappa}{2}. 
  \end{array}
\end{equation}
Note that now, in contrast to the regular lattice case of
Eqs.~(\ref{eq:critical_operators}) and (\ref{eq:tricritical_operators}), all weights
depend {\em linearly\/} on the parameter $\kappa$. The resulting FK and geometric
cluster fractal dimensions for the Potts model coupled to dynamical triangulations
are summarised in Table \ref{tab:exponents}.

\begin{table}[tb]
  \renewcommand{\arraystretch}{1.5}
  \caption{Cluster fractal dimensions of the $q$-state Potts model on dynamical
    triangulations. The values of $\td_\mathrm{EP}/d_h$ and $\td_\mathrm{RB}/d_h$ for the geometric clusters
    are only formal since, by definition, $\td_\mathrm{EP}/d_h \le
    \td_\mathrm{H}/d_h$. Thus one has $\td_\mathrm{EP}^\mathrm{G}/d_h =
    \td_\mathrm{H}^\mathrm{G}/d_h$ and the geometric clusters are multiply connected.} 
  \centering
  \vspace*{2ex}
  \begin{tabular}{cc|ccccc} \hline\hline
    $q$ & Type & $\td_\mathrm{C}/d_h$ & $\td_\mathrm{H}/d_h$ & $\td_\mathrm{EP}/d_h$
    & $\td_\mathrm{RB}/d_h$ \\ \hline
      & FK   & $\frac{7}{8}$   & $\frac{3}{4}$ & $\frac{1}{2}$ & $\frac{1}{4}$ \\
    \raisebox{2ex}[2ex]{1} & Geo  & $1$             & $\frac{1}{2}$ & $(\frac{3}{4})$ & $(-\frac{1}{4})$ \\ \hline
      & FK   & $\frac{5}{6}$   & $\frac{2}{3}$ & $\frac{1}{2}$ & $\frac{1}{6}$ \\
    \raisebox{2ex}[2ex]{2} & Geo  & $\frac{11}{12}$ & $\frac{1}{2}$ & $(\frac{2}{3})$ & $(-\frac{1}{6})$ \\ \hline
      & FK   & $\frac{4}{5}$   & $\frac{3}{5}$ & $\frac{1}{2}$ & $\frac{1}{10}$ \\
    \raisebox{2ex}[2ex]{3} & Geo  & $\frac{17}{20}$ & $\frac{1}{2}$ & $(\frac{3}{5})$ & $(-\frac{1}{10})$ \\ \hline
      & FK   & $\frac{3}{4}$   & $\frac{1}{2}$ & $\frac{1}{2}$ & $0$ \\
    \raisebox{2ex}[2ex]{4} & Geo  & $\frac{3}{4}$   & $\frac{1}{2}$ & $(\frac{1}{2})$ & $0$ \\\hline\hline
  \end{tabular}
  \label{tab:exponents}
\end{table}

\section{Conclusions}

We have shown how the fractal dimensions associated with the geometric clusters of
the Potts model coupled to the dynamical triangulations model of Euclidean quantum
gravity in two dimensions can be inferred from a mapping to the tricritical branch of
the Potts model. To allow for an application of the KPZ framework, the conformal
weights associated to the fractal dimensions $d_\mathrm{C}$ of the cluster,
$d_\mathrm{H}$ of the cluster hull, $d_\mathrm{EP}$ of the external perimeter and
$d_\mathrm{RB}$ of the cluster red bonds have been identified from the Kac table.
Lifting the corresponding weights of the critical and tricritical branches to the
dynamical triangulations results in rational values for the fractal dimensions of FK
and geometric clusters for the $q=0$, $1$, $2$, $3$ and $4$ state Potts models
summarised in Table \ref{tab:exponents}. The duality symmetry $\kappa \rightarrow
1/\kappa$ present between the critical and tricritical weights on regular lattices,
Eqs.~(\ref{eq:critical_operators}) and (\ref{eq:tricritical_operators}), is no longer
present in the dressed weights of Eqs.~(\ref{eq:dressed_critical}) and
(\ref{eq:dressed_tricritical}). As a peculiarity, we note that the fractal dimension
of the geometric cluster hulls is a constant $\td_\mathrm{H}^\mathrm{G}/d_h=1/2$
independent of $q$. A scaling or duality relation between the dimensions of the
external perimeter and cluster hulls found valid for the regular case
\cite{duplantier:00a},
\begin{equation}
  \label{eq:duplantier}
  (d_\mathrm{EP}/d-\frac{1}{2})(d_\mathrm{H}/d-\frac{1}{2})= \frac{1}{16},
\end{equation}
does not apply to the model coupled to quantum gravity, but instead one gets,
\begin{equation}
  \label{eq:duplantier_dressed}
  (\td_\mathrm{EP}/d_h-\frac{1}{2})(\td_\mathrm{H}/d_h-\frac{1}{2})=0.
\end{equation}
Also, while the four fractal dimensions fulfil the identity
$d_\mathrm{C}-d_\mathrm{H}=\frac{1}{4}(d_\mathrm{EP}-d_\mathrm{RB})$ for regular
lattices \cite{janke}, for the random-lattice case we instead find
$\td_\mathrm{C}-\td_\mathrm{H}=\frac{1}{2}(\td_\mathrm{EP}-\td_\mathrm{RB})$. A
couple of further similar scaling relations can be formulated.

Our simulation results for the case of the $q=2$ Potts or Ising model coupled to
dynamical triangulations show full consistency of the measured fractal dimensions of
the Fortuin-Kasteleyn and geometric clusters with the predictions resulting from
the KPZ mapping, confirming that the properties of the geometric Potts clusters on
dynamical triangulations are described by the corresponding tricritical Potts model
of the same central charge. Corrections to scaling are found to be much stronger for
the random-graph model than for the square-lattice simulations performed as a gauge.
These corrections, known to result from the small effective linear extents of the
graphs due to their large Hausdorff dimension $d_h \approx 4$ \cite{weigel:04b}, have
to be explicitly taken into account to find consistency with the scaling predictions
and satisfactory quality of the fits.

It would be interesting to see whether, as expected, the fractal dimensions of the
geometric clusters of the $q \ne 2$ Potts models coupled to dynamical
triangulations follow the predictions summarised in Table \ref{tab:exponents}, in
particular for the case of $q=4$, where the critical and tricritical branches
coalesce and the KPZ mapping (\ref{dressing_KPZ}) becomes marginal due to central
charge $c=1$.

\section*{Acknowledgements}

This work was partially supported by the EC RTN-Network `ENRAGE': {\em Random
  Geometry and Random Matrices: From Quantum Gravity to Econophysics\/} under grant
No.~MRTN-CT-2004-005616. M.W.\ acknowledges support by the EC ``Marie Curie
Individual Intra-European Fellowships'' programme under contract No.\
MEIF-CT-2004-501422.

%\bibliography{general}

\end{document}